 \documentclass[doublecol]{epl2} 

\usepackage{amssymb}
\usepackage{amsmath}
\usepackage{gensymb}
\usepackage{graphicx}
\usepackage{mathtools}
\usepackage{epstopdf}
\usepackage{epsfig}
\usepackage{float}

\newcommand\PT{\mathcal{PT}}

\title{Exactly solvable $\PT$-symmetric models in two dimensions}

\author{Kaustubh S. Agarwal\inst{1} \and Rajeev K. Pathak\inst{1} \and Yogesh N. Joglekar\inst{2}}
\shortauthor{Agarwal \etal}

\institute
{ \inst{1} Department of Physics, Savitribai Phule Pune University, Ganeshkhind, Pune 411007, Maharashtra, India\\
  \inst{2} Department of Physics, Indiana University-Purdue University Indianapolis (IUPUI), Indianapolis, Indiana 46202, USA
}


\abstract{Non-hermitian, $\PT$-symmetric Hamiltonians, experimentally realized in optical systems, accurately model the properties of open, bosonic systems with balanced, spatially separated gain and loss. We present a family of exactly solvable, two-dimensional, $\PT$ potentials for a non-relativistic particle confined in a circular geometry. We show that the $\PT$ symmetry threshold can be tuned by introducing a second gain-loss potential or its hermitian counterpart. Our results explicitly demonstrate that $\PT$ breaking in two dimensions has a rich phase diagram, with multiple re-entrant $\PT$ symmetric phases. 
}

\begin{document}
\maketitle


\section{Introduction} In their pioneering paper, Bender and Boettcher~\cite{cbsb1998} demonstrated that the conventional hermiticity requirement for a quantum Hamiltonian is rather restrictive. While it is sufficient to engender real eigenvalues and eigenvectors that are orthonormal with respect to the standard inner product, they showed that a large class of continuum Hamiltonians on an infinite line that are invariant under the composite operation of parity ($\mathcal{P}$) and time-reversal ($\mathcal{T}$) have a purely real spectrum, albeit with non-orthogonal eigenfunctions. Bender {\it et al.} showed that in such cases, the eigenfunctions are orthogonal with respect to a new, Hamiltonian-dependent inner product, and thus, a unitary, self-consistent complex extension of quantum mechanics can be developed for $\PT$ Hamiltonians in the parameter region where the eigenvalues are purely real~\cite{cb2001}. Note that in this approach, only operators that are self-adjoint under the new inner product are observables, and due to the Hamiltonian-dependent nature of the inner product, generically, the set of observables contains only the Hamiltonian. Subsequently, Mostafazadeh established that a $\PT$-symmetric, non-Hermitian Hamiltonian with purely real spectrum can be transformed into a Hermitian Hamiltonian under a \emph{similarity} transformation~\cite{am2002}, instead of the usual unitary transformation, and therefore such Hamiltonians are aptly termed \emph{pseudo-Hermitian} Hamiltonians~\cite{am2010}. 

Since then, several authors have investigated properties of one dimensional, exactly solvable, non-Hermitian, $\PT$-symmetric Hamiltonians with a finite support~\cite{ex1,ex3,ex4,ex5,ex6}, although such examples are scarce. These studies are complemented by those of several continuum~\cite{cm1hj2012,m1,m2,m3,m4} and discrete~\cite{m5,m8,m9,m10,m11,m12,m13,m14} $\PT$ symmetric Hamiltonians. A $\PT$ Hamiltonian, continuum or discrete, typically consists of a Hermitian kinetic energy term $H_0$ and complex, $\PT$-symmetric potential term $V\neq V^\dagger$. When the strength of the complex potential is small, the spectrum of the $\PT$ Hamiltonian $H=H_0+V$ is purely real. It evolves into complex-conjugate pairs when the strength of the potential exceeds a threshold that is determined by the energy scale of the Hermitian part. At the $\PT$-symmetry breaking threshold, where the spectrum changes from real to complex, the Hamiltonian $H$ is defective, owing to the fact that a pair of eigenvalues of $H$, as well as corresponding eigenvectors become degenerate at the exceptional point~\cite{kato,heiss,rotter,yjreview}. 

During the past five years, it has become clear that $\PT$ Hamiltonians, while perhaps not fundamental in nature, faithfully describe open systems with balanced gain and loss. We note that, in contrast with the initial work on $\PT$-symmetric Hamiltonians~\cite{cb2007}, this {\it effective} Hamiltonian description requires the use of standard inner product and the non-unitary time evolution generated by the Hamiltonian is understood in terms of the interaction of the open system with the environment. The notion of amplification and decay of a {\it single} quantum state is applicable to bosons, and the implementation of gain and loss is, most easily, implemented for photons. Thus, the $\PT$ breaking transition and effects of the exceptional point have been experimentally observed in coupled optical waveguides~\cite{e1,e2,e3}, micro-resonators~\cite{e4,e5}, lasers~\cite{e6,e7}, and two-dimensional (2D) honeycomb photonic crystals~\cite{e75,e8}, as well as in coupled electrical~\cite{e9} and mechanical oscillators~\cite{e10}. In addition, it is now apparent that complex potentials arise in the effective action for a wide range of non-equilibrium condensed matter systems including type-II superconductors in transverse field~\cite{hatano}, driven superconducting wires~\cite{ma2007}, voltage-biased superconducting junction~\cite{serbyn2013}, and Coulomb gases~\cite{gulden2013}. 

All of the studies mentioned above, with the single exception~\cite{e75,e8}, are confined to one dimensional problems with a non-degenerate spectrum for the Hermitian kinetic energy term $H_0$. Perturbatively, this absence of degeneracy implies that the $\PT$-symmetry breaking threshold is positive~\cite{moiseyev}.  In two or higher dimensions, the spectrum of the $H_0$ is degenerate for both lattice and continuum models. A generic $\PT$-symmetric potential $V$ connects these degenerate states and thus leads to a complex energy spectrum at arbitrarily small strengths of $V$~\cite{cmb2013,mandal2013}. A group theoretical analysis of these degeneracies and their effect on the $\PT$-breaking transition threshold was carried out by Ge and Stone~\cite{gestone2014}. L\'evai proposed the general formalism for treating two and three dimensional  $\PT$-symmetric potentials~\cite{gl2008}. 

In this paper, we present a family of 2D complex potentials with analytically tractable $\PT$-symmetry breaking threshold. In the next section, we obtain the $\PT$ phase diagram of a non-relativistic particle confined to an annulus of arbitrary inner and outer radii. The $\PT$ phase diagram in the presence of one or more such potentials is discussed in the subsequent section. Lastly we show that adding a Hermitian counterpart to a balanced gain-loss potential leads to a rich variety of $\PT$ phases. Our results provide several examples of analytically solvable, two-dimensional, continuum $\PT$ Hamiltonians, and demonstrate the consequences of the competition among level attraction induced by non-Hermitian gain-loss potentials, and level repulsion induced by Hermitian potentials. 


\section{Model}
\label{sec:model}
Let us consider a particle of mass $\mu$ confined to a circular region of inner radius $a_<$ and outer radius $a_>$. The eigenvalue problem for such a particle in the presence of a time-independent $\PT$-symmetric potential $V(\rho,\phi)$ is given by 
\begin{equation}
\label{eq.1}
-\frac{\hbar^2}{2\mu}\nabla^2\Psi_q(\rho,\phi) + V(\rho,\phi)\Psi_q(\rho,\phi) = E_q\Psi_q(\rho,\phi), 
\end{equation}
where $(\rho,\phi)$ are the cylindrical two-dimensional coordinates, $V(\rho,\phi)$ is the complex, non-Hermitian potential, and $q$ denotes a complete set of quantum numbers that uniquely specify an eigenfunction $\Psi_q$ with energy $E_q$. The eigenfunctions, by construction, have a finite support and satisfy boundary condition $\Psi_q(\rho= a_<,\phi)=0=\Psi_q(\rho=a_>,\phi)$. 

Consider the following family of pure gain-loss potentials with strength $\gamma>0$, 
\begin{equation}
\label{Potential}
V_n(\rho,\phi)=-i\gamma_{n}\cos(n\phi)/\rho^2,
\end{equation}
where $n$ is an odd integer, and $V_n$ is nonzero only inside the region $a_<\leq\rho\leq a_>$. Thus 
the regions with $\cos(n\phi)<0$ are the gain regions and those with $\cos(n\phi)>0$ are the lossy regions. This potential is odd under parity and is symmetric under $\PT$ transformation where the parity operator is given by $\mathcal{P}:(\rho,\phi)\rightarrow (\rho,\phi+\pi)$ and the time-reversal operator is given by $\mathcal{T}:i\rightarrow -i$. In the limit $a_<=0$, the potential $V_n$ diverges at the origin. Nonetheless, as we will show in the following paragraphs, it has a positive $\PT$-symmetry breaking threshold~\cite{yda2014}.  

 The $1/\rho^2$ dependence of the gain-loss potential allows us to decouple the problem into angular and radial sectors $\Psi(\rho,\phi) = R(\rho)\Phi(\phi)$, and obtain uncoupled differential equations for each sector, 
\begin{eqnarray}
\label{eq:radial}
\rho^2\partial_\rho^2 R(\rho) + \rho\partial_\rho R(\rho)+(\rho^2\kappa^2 -\alpha^2)R(\rho) & = & 0,\\
\label{eq:polar}
\partial_\phi^2\Phi(\phi) + i\beta_n \cos(n\phi)\Phi(\phi)+\alpha^2\Phi(\phi) & = & 0.
\end{eqnarray}
where $\beta_{n}=\gamma_n/(\hbar^2/2\mu)\geq 0$ is the dimensionless strength of the gain-loss potential, $\kappa^2=2\mu E/\hbar^2$, and the radial wavefunction obeys boundary conditions $R(\rho=a_<)=0$ and $R(\rho=a_>)=0$. Since $\beta_n$ is independent of the length-scales in the problem, we expect that the $\mathcal{PT}$-breaking threshold will be independent of the domain size. It follows from eqs.(\ref{eq:radial})-(\ref{eq:polar}) that the eigenvalues of the 2D Hamiltonian $H(\beta_n)=H_0+V_n$ are real if and only if $\alpha^2$ is real. Therefore, the $\PT$ breaking threshold $\beta_{nc}$ for a given potential $V_n(\rho,\phi)$ is obtained by analyzing the spectrum of eq.(\ref{eq:polar}) and is independent of the radial sector equation. 

In the angular momentum basis $|m\rangle$, where $\langle\phi|m\rangle=\exp(im\phi)$, the polar sector equation translates into an eigenvalue equation for a tridiagonal, non-Hermitian, symmetric matrix
\begin{equation}
\label{eq:A}
A_{mm'}=m^2\delta_{mm'} -\frac{i\beta_n}{2}(\delta_{m,m'+n}+\delta_{m,m'-n})=A_{m'm}
\end{equation}
where $m,m'\in\mathbb{Z}$. When the $\PT$ potential is zero, the spectrum of $A$ is trivially given by $\alpha^2=m^2$ and is doubly-degenerate for all $m$ except $m=0$. As the strength of the non-Hermitian potential $\beta_n$ is increased, a pair of consecutive $\alpha^2$ eigenvalues become degenerate and then complex, thus defining the $\PT$ breaking threshold $\beta_{nc}$. In practice, for numerical calculations, we truncate the infinite-dimensional matrix $A$ to a $(2M+1)\times(2M+1)$ matrix and choose the angular momentum cutoff $|m|\leq M\sim 100$ such that the results are independent of it.  

\begin{figure}
\centering
\includegraphics[width=\columnwidth]{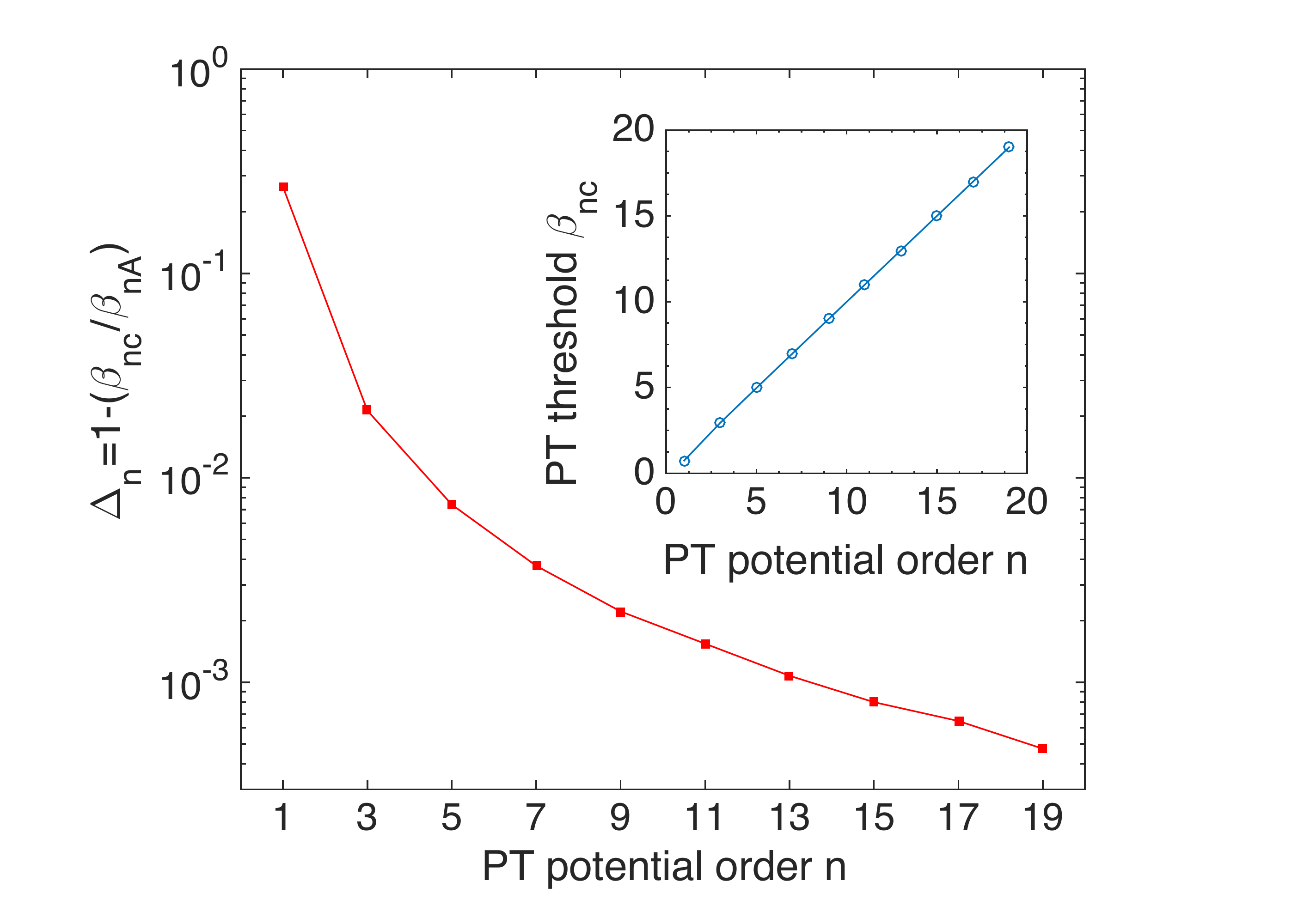}
\caption{Comparison of $\PT$ breaking thresholds for $V_n$ obtained analytically, $\beta_{nA}=n$ and numerically, $\beta_{nc}$. The inset shows that $\beta_{nc}$ is equal to $n$ except at small values of $n$. The main panel shows that the error $\Delta_n$ decays exponentially with $n$.}
\label{fig:betac}
\end{figure}
It is possible to obtain the $\PT$-breaking threshold {\it analytically} by truncating the matrix $A$ to only angular momentum levels involved in the $\PT$ symmetry breaking. For example, when $n=3$, the levels $\alpha^2=\{1,4\}$ become degenerate, and correspond to angular momentum levels $(m,m')=(\pm1,\mp 2)$; when $n=5$, the levels that become degenerate are $\alpha^2=\{4,9\}$ or $(m,m')=(\pm2,\mp3)$. Thus for a given $n$, the consecutive levels that become degenerate are $\alpha^2=\{(n-1)^2/4,(n+1)^2/4\}$ and correspond to angular momenta $m=\pm(n-1)/2$ and $m'=\mp(n+1)/2$ that differ by $n$. Therefore the matrix $A$ becomes $A_2=(n^2+1)/2 -(n/2)\sigma_z -(i\beta_n/2)\sigma_x=A_2^T\neq A_2^\dagger$, where $\sigma_i$ are the usual Pauli matrices. It then follows that {\it the analytical threshold for the potential $V_n$ is given by $\beta_{nA}=n$} or, equivalently, 
\begin{equation}
\label{eq:threshold}
\gamma_{nA}=n\frac{\hbar^2}{2\mu}.
\end{equation}
The inset in fig.~\ref{fig:betac} shows that the numerically obtained $\PT$-breaking threshold $\beta_{nc}$ is equal to $n$ except at small $n$.  The main panel in fig.~\ref{fig:betac} shows that the difference between the analytical prediction $\beta_{nA}=n$ and the exact, numerical result $\beta_{nc}$ decays exponentially with the order $n$ of the potential. This exponentially decaying correction to the $2\times2$ approximation can be explained by Salwen's perturbation theory~\cite{salwen,tonylee}. When $n=1$, a better estimate for the threshold $\beta_{nA}$ is obtained by truncating the matrix $A$ to its $(m,m')=\{-1,0,1\}$ sector. It gives $\beta_{1A}=1/\sqrt{2}=0.707$, instead of a threshold strength of unity as predicted by the earlier analysis, and compares more favorably with the exact, numerical result $\beta_{1c}=0.7350$. 

\begin{figure}
\centering
\includegraphics[width=1\columnwidth]{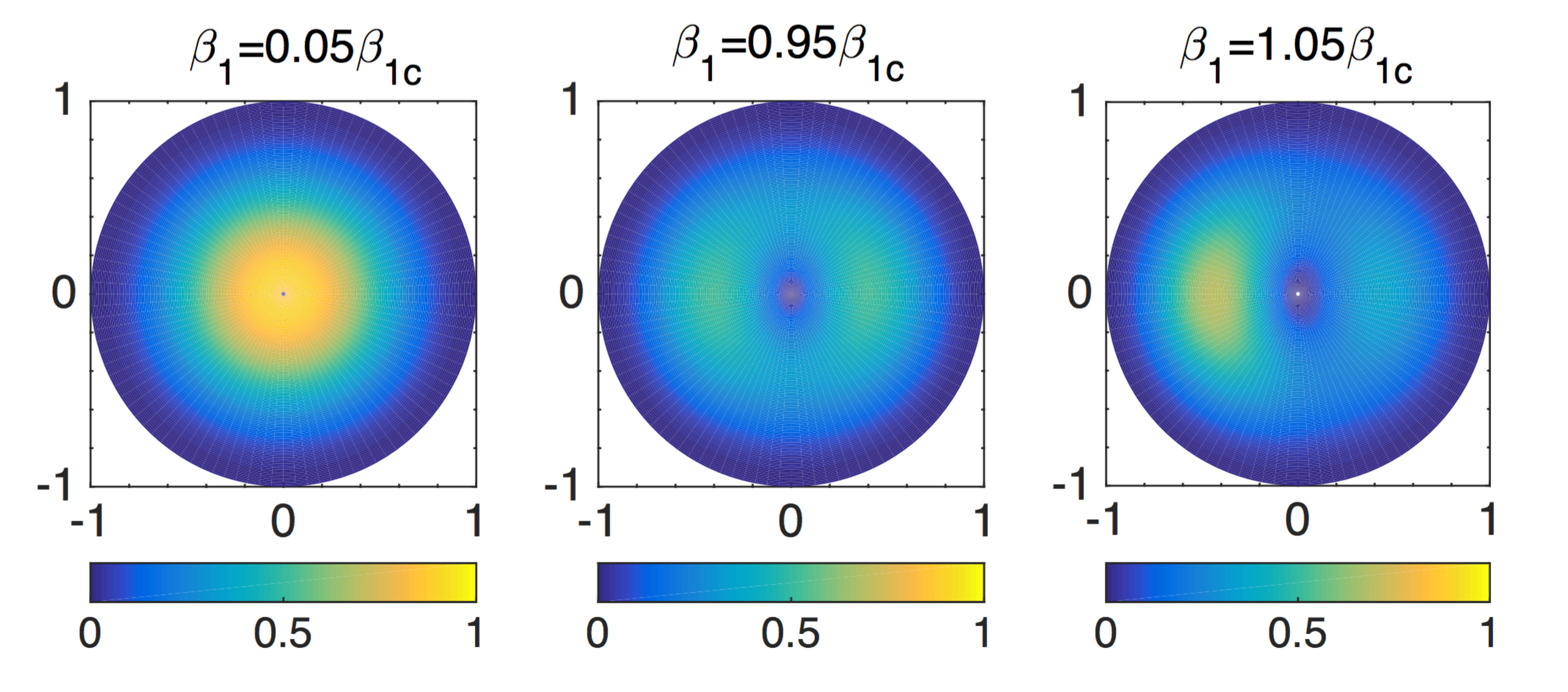}
\caption{Ground state probability density for potential $V_1(\rho,\phi)$ in the x-y plane; both axes are scaled by the domain radius $a_>$. The probability is parity symmetric when $\beta_1\leq\beta_{1c}$, whereas in the $\PT$-broken region, $\beta_1>\beta_{1c}$, the ground state wavefunction $\Psi_G(\rho,\phi)$ has a higher weight in the gain region.}
\label{fig:WfC2}
\end{figure}
 Although the polar sector, eq.(\ref{eq:polar}), is sufficient to determine the $\PT$-breaking threshold, it is the solution of the radial equation, $R(\rho)=c_1 J_\alpha(\kappa\rho) +c_2 Y_\alpha(\kappa\rho)$, that determines the energy spectrum. In particular, for a particle confined in a cylinder, the quantized values of $\kappa_q$ are determined by the zeros of the Bessel function $J_\alpha(\kappa_q a_>)=0$ and lead to the corresponding spectrum $E(\alpha,q)=\hbar^2\kappa_q^2/2\mu$. In the $\mathcal{PT}$-symmetric phase, the eigenfunctions $\Psi(\rho,\phi)$ are $\PT$ symmetric, and therefore their probability density is invariant under rotation by $\pi$; in the $\mathcal{PT}$ broken phase, this symmetry disappears, and the eigenfunctions have a higher weight in the gain region. These generic features are illustrated in fig.~\ref{fig:WfC2}, which shows the evolution of the ground state probability density for potential $V_1(\rho,\phi)=-i\gamma_1\cos(\phi)/\rho^2$ at three different values of $\beta=\gamma/(\hbar^2/2\mu)$. At small non-Hermiticity, $\beta_1/\beta_{1c}=0.05$, the ground state wavefunction is close to that of the Hermitian problem, $\Psi_G(\rho,\phi)=J_0(\kappa\rho)$ where $\kappa a_>= 2.4048$ corresponds to the first zero of the Bessel function. Therefore, the probability distribution is isotropic, and has no nodes except at the boundary (left-hand panel). As $\beta_1$ is increased, due to the increased transitions between the $m=0$ and $m=\pm 1$ angular momentum states, the probability distribution shifts away from the origin, develops angular structure, but retains symmetry under the 2D parity operation (center panel). The right-hand panel shows that as the $\PT$ potential strength exceeds the threshold, $\beta_1=1.05\beta_{1c}$, the parameter $\alpha$ becomes complex and the resulting ground state has higher probability in the gain region $\pi/2<\phi<3\pi/2$. 

When the order $\alpha$ of the Bessel function becomes a complex number, the boundary constraint $J_\alpha(\kappa_qa_>)=0$ is only satisfied for complex values of $\kappa_q$ which, in turn, lead to an infinite sequence of complex eigenenergies $E(\alpha,q)$. This behavior contrasts with that of one dimensional models on a finite segment, where only a finite set of energy levels participate in the $\PT$ symmetry breaking (with a few exceptions~\cite{m10,jake}). In both cases, however, the set of $\PT$ symmetry breaking levels has co-dimension one. 


\section{Phase diagram with two $\PT$ potentials} 
\label{sec:ptc}
In the previous section, we considered the $\PT$ symmetry breaking threshold in the presence of a single, purely imaginary potential $V_n$ where the sign of the potential did not matter. In this section, we will investigate the $\PT$ threshold in the presence of two such potentials. The Hamiltonian for such a system is given by $H(\beta_n,\beta_m)=H_0+V_n+V_m$ where $n,m$ are odd integers. Since the spectrum of $H$ is either purely real or has complex conjugate pairs, it is identical to that of $H^*$ where $*$ denotes the complex conjugation operation. It follows that the $\PT$ breaking threshold is the same for $H(\beta_n,\beta_m)$ and $H(-\beta_n,-\beta_m)$, but the thresholds for $H(\beta_n,\beta_m)$ and $H(-\beta_n,\beta_m)$ are, in general, different. 

\begin{figure}
\hspace{-5mm}
\includegraphics[width=\columnwidth]{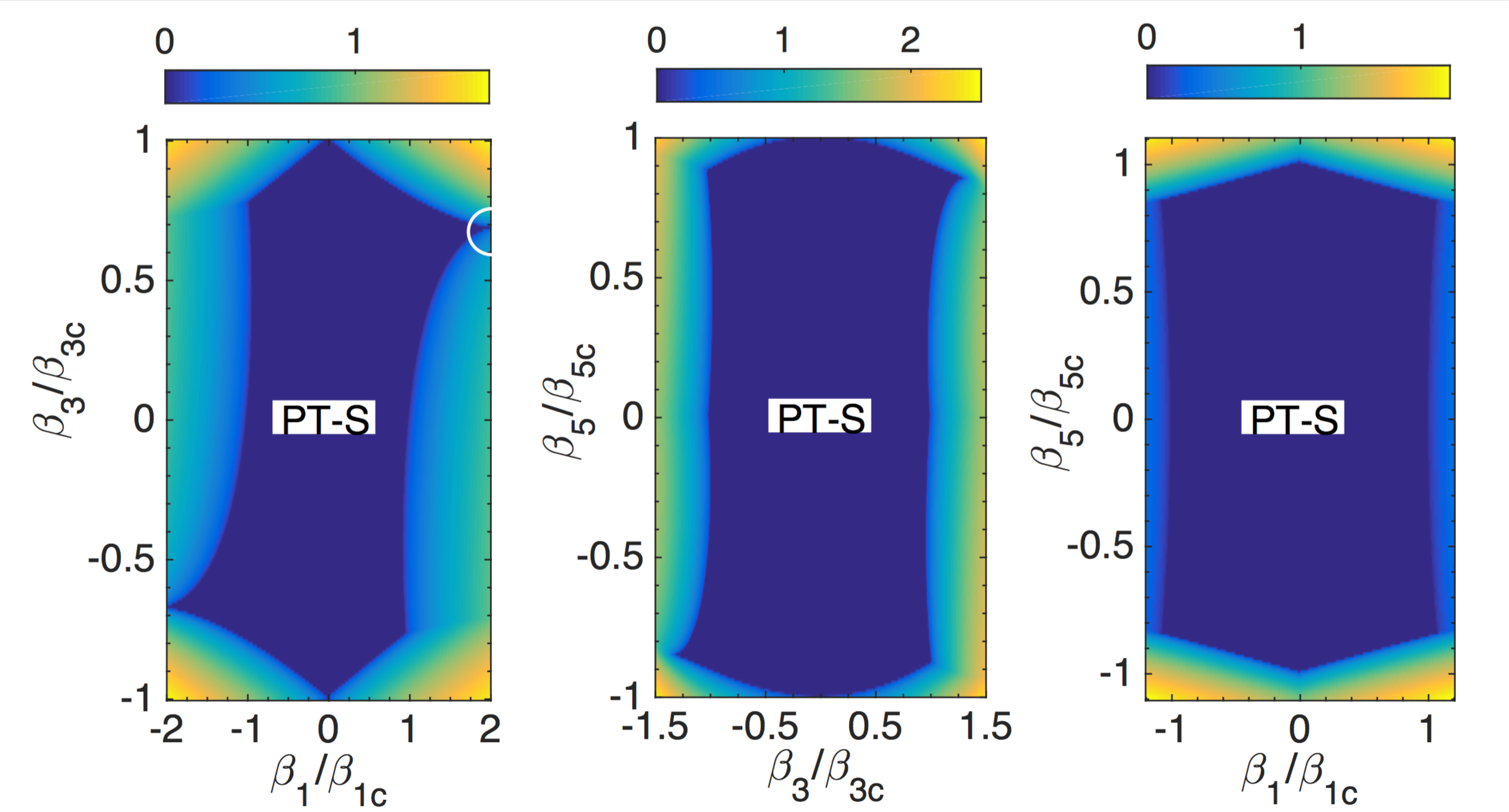}
\caption{$\PT$ phase diagram for competing gain-loss potentials. Each potential strength $\beta_k$ is normalized to its threshold $\beta_{kc}$, the color denotes the largest imaginary part of the spectrum of $H(\beta_n,\beta_m)$, and deep blue denotes the $\PT$-symmetric (PT-S) phase. The left-hand and center panels show the enhancement of $\PT$ threshold. The right-hand panel shows that $\PT$ threshold is either suppressed or unaffected when the two gain-loss potentials do not act on a common energy level.}
\label{fig:ptphases}
\end{figure}
Figure~\ref{fig:ptphases} shows the numerically obtained $\PT$ phase diagrams for Hamiltonian $H(\beta_n,\beta_m)$ in the $(\beta_n,\beta_m)$ plane where each axis is scaled by the corresponding $\PT$-breaking threshold strength; plotted in each panel is the largest imaginary part of the spectrum of $H$ and the $\PT$-symmetric region (PT-S) is shown in deep blue. With this scaling of the axes, for uncorrelated potentials $V_n$ and $V_m$, we expect the $\PT$-symmetric phase to be confined to the central square region $-1\leq\beta_n/\beta_{nc}\leq 1$ and $-1\leq\beta_m/\beta_{mc}\leq 1$.

The left-hand panel in fig.~\ref{fig:ptphases} shows that for $H(\beta_1,\beta_3)$ as the potential $\beta_3$ is increased, the $\PT$ symmetric phase extends well beyond the square and doubles in width to $|\beta_1|\sim 2\beta_{1c}$ near $\beta_3/\beta_{3c}=0.7$, marked by a white circle. This increase occurs only in the first and fourth quadrant, when the two potentials $V_1$ and $V_3$ have largely overlapping gain regions. This strengthening of the $\PT$ breaking threshold also leads to re-entrant $\PT$ symmetric phase, where increasing the gain-loss magnitude $\beta_3$ restores a broken $\PT$ symmetry~\cite{yjbb2012}. The central panel shows a similar enhancement of $\PT$ symmetric phase beyond the expected square region for Hamiltonian $H(\beta_3,\beta_5)$, although, in the present case, the maximum enhancement is about 40\%. The right-hand panel in fig.~\ref{fig:ptphases} shows the resulting phase diagram for $H(\beta_1,\beta_5)$. In contrast to the first two cases, here, we see that the $\PT$ symmetric phase is not enhanced. Instead, it is confined to the central square region, with further suppression near its four corners. Incidentally, we note that all numerically obtained phase diagrams are symmetric under reflection through the origin $(\beta_n,\beta_m)\rightarrow -(\beta_n,\beta_m)$. 

\begin{figure}[h!]
\centering
\includegraphics[width=\columnwidth]{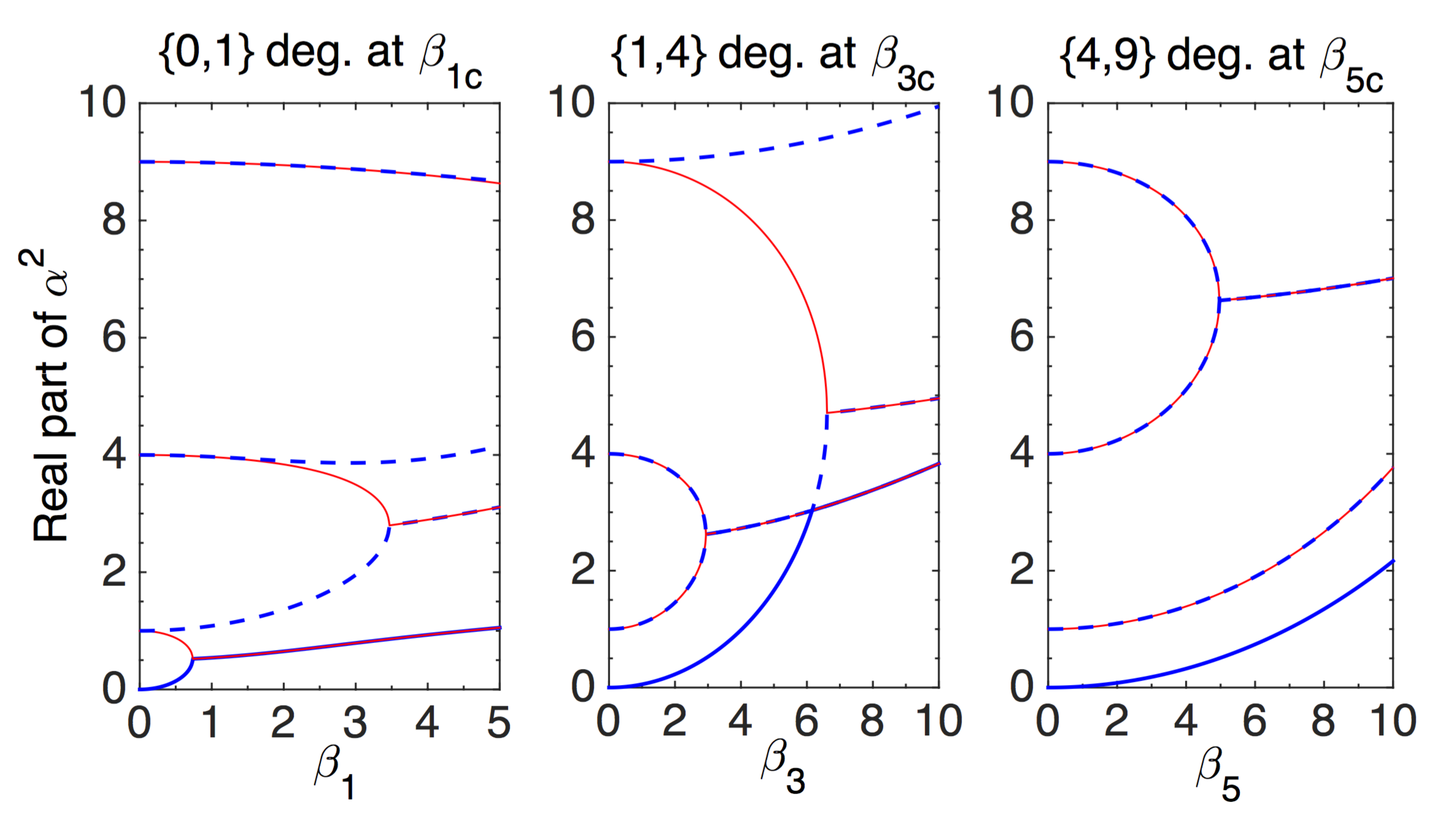}
\caption{Flow of first four eigenvalues $\alpha^2(\beta)$ of matrix $A$ for potentials $V_1$ (left-hand panel), $V_3$ (center panel), and $V_5$ (right-hand panel). Adjacent panels share a single eigenvalue that is pushed down in the first panel and up in the second, thus leading to enhancement of the $\PT$ breaking threshold for a system with both potentials present.}
\label{fig:ptflow}
\end{figure}

The surprising results in fig.~\ref{fig:ptphases}, which show both substantial enhancement or minor suppression of the $\PT$ breaking threshold, are understood most easily by the action of a single gain-loss potential $V_k$ on the $\alpha^2$-levels that participate in the $\PT$-symmetry breaking (fig.~\ref{fig:ptflow}). Each panel in fig.~\ref{fig:ptflow} shows the flow of first four eigenvalues of the matrix $A$, eq.(\ref{eq:A}), as a function of the potential strength $\beta_k$. The potential $V_1$ breaks $\PT$ symmetry by pushing down the $\alpha^2=1$ level and raising the $\alpha^2=0$ level so that the two become degenerate at the threshold $\beta_{1c}=0.7350$ (left-hand panel). Note that a second, linearly independent eigenstate in the $\alpha^2=1$ level becomes degenerate with the $\alpha^2=4$ level at a higher value of gain-loss strength $\beta_1\sim 3.5$; however, the $\PT$-breaking threshold is determined by the first occurrence of such a degeneracy. Similarly, potential $V_3$ breaks the $\PT$ symmetry by lowering the $\alpha^2=4$ level and pushing up the $\alpha^2=1$ level (center panel), and the potential $V_5$ lowers the $\alpha^2=9$ level and pushes up the $\alpha^2=4$ level (right-hand panel). Therefore, potentials $V_1$ and $V_3$ have competing, opposite effect on level $\alpha^2=1$, as do potentials $V_3$ and $V_5$ on the level $\alpha^2=4$. {\it This competition is the cause of the enlargement of $\PT$ symmetric phase} in the two cases. In contrast, potentials $V_1$ and $V_5$ do not share a single eigenlevel that participates in both $\PT$ breaking transitions. Therefore, the $\PT$ symmetric phase is confined to the central square region (fig.~\ref{fig:ptphases}). 


\section{Phase diagram with Hermitian and $\PT$ potentials}
\label{sec:pth} Let us now consider the effect of adding a Hermitian potential $U_p$ to the original Hamiltonian, $H=H_0+V_n$. We choose a family of potentials 
\begin{equation}
\label{eq:upot}
U_p(\rho,\phi)=-l_p\cos(p\phi)/\rho^2,
\end{equation}
where $p$ is an even integer and the dimensionless strength of the potential is given by $\lambda_p=l_p/(\hbar^2/2\mu)$. An even $p$ ensures that the total Hamiltonian $H(\lambda_p,\beta_n)=H_0+U_p+V_n$ is also $\PT$ symmetric, and therefore has a spectrum that is invariant under complex conjugation.  It follows then that the $\PT$-symmetric phase diagram is invariant under $\beta_n\rightarrow-\beta_n$; physically this corresponds to exchanging the gain and loss regions. 
Note that, in general, no such symmetry can be assigned to  $\lambda_p\rightarrow-\lambda_p$, which corresponds to exchanging the attractive-potential-regions, which might support bound states, with repulsive-potential-regions. 

\begin{figure}
\centering
\includegraphics[width=\columnwidth]{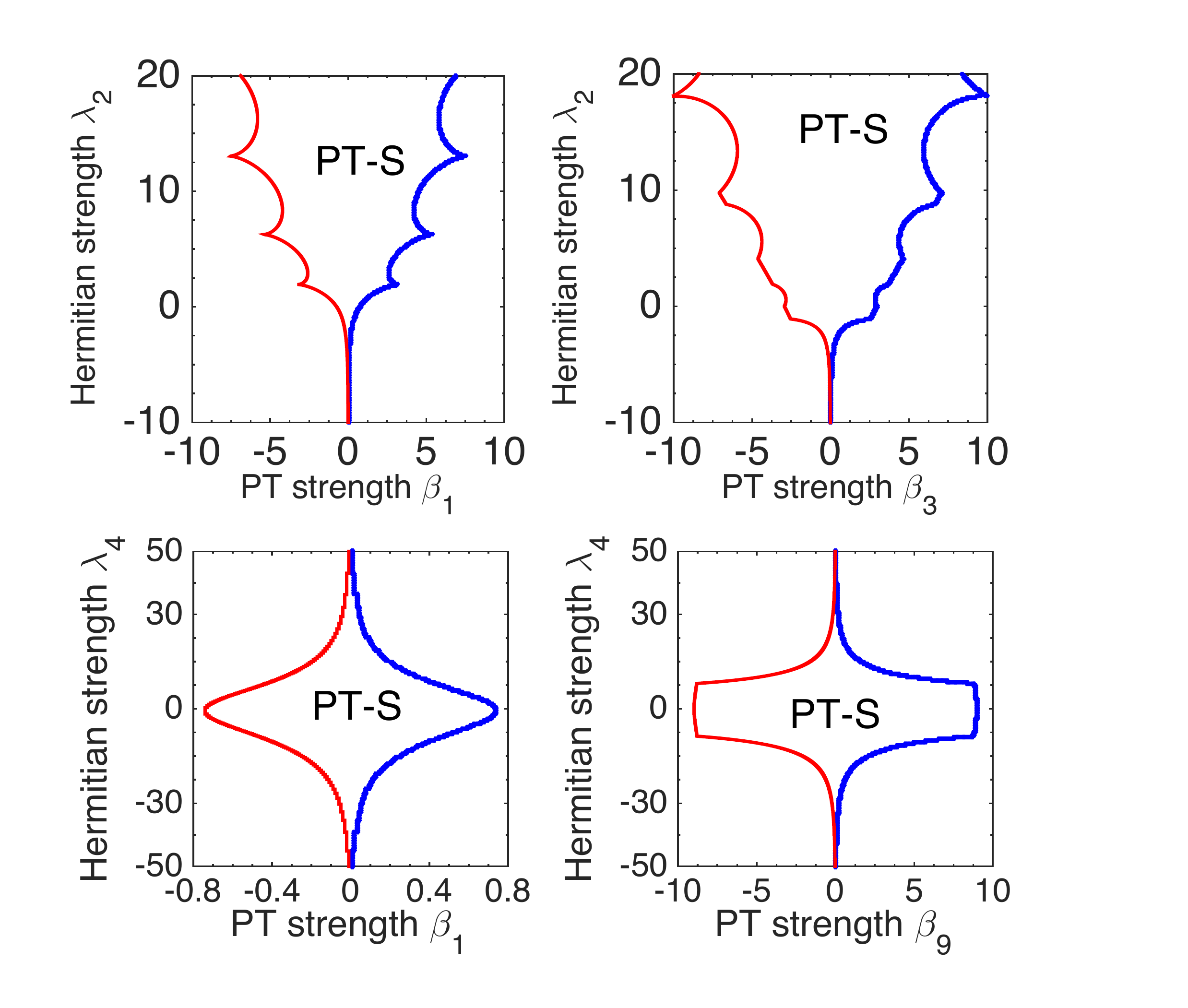}
\caption{Top row: phase diagrams with Hermitian potential $U_2$ show that the $\PT$ threshold $\beta_{nc}(\lambda_2)$ is asymmetrically enhanced or suppressed based on the sign of $\lambda_2$. Bottom row: phase diagrams with potential $U_4$ show that $\PT$ threshold $\beta_{nc}(\lambda_4)$ is symmetrically suppressed when $\lambda_4\neq0$. These results are true for gain-loss potential with any (odd) order $n$.}
\label{fig:nhh24}
\end{figure}
\begin{figure}[h]
\centering
\includegraphics[width=\columnwidth]{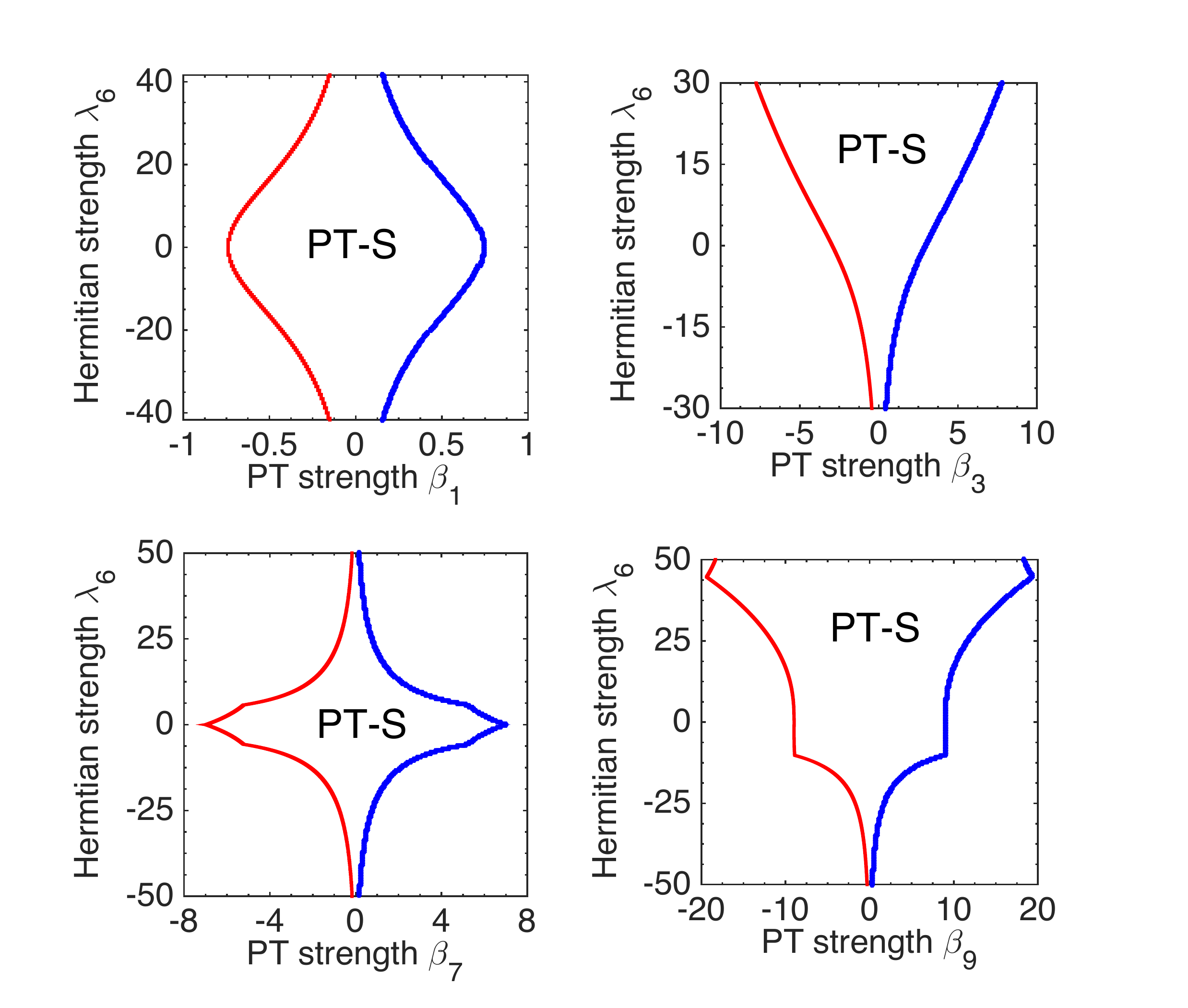}
\caption{$\PT$ phase diagram for the Hamiltonian $H(\lambda_p,\beta_n)$ with $p=6$ and different orders of the gain-loss potential. The phase diagram is asymmetric in $\lambda_6$ only when $n$ is a multiple of $p/2=3$; for all other $n$, it shows an unexpected symmetry under $\lambda_6\leftrightarrow-\lambda_6$.}
\label{fig:nhh6}
\end{figure}

Figure~\ref{fig:nhh24} shows typical phase diagrams for the Hamiltonian $H(\lambda_p,\beta_n)$ with $p=2$ (top row) and $p=4$ (bottom row). The blue and red phase boundaries, obtained numerically for $\beta_n\geq0$ and $\beta_n\leq0$ respectively, confirm the phase-diagram symmetry. The horizontal axis in each panel denotes the absolute dimensionless gain-loss strength $\beta_n$; thus, the $\PT$-breaking threshold when $\lambda_p=0$ is approximately given by $\beta_{nc}\approx\beta_{nA}=n$. The top row shows that for potential $U_2$, the $\PT$ threshold $\beta_{nc}(\lambda_2)$ increases roughly linearly with $\lambda_2\geq 0$, but is suppressed for $\lambda_2<0$, thus resulting in a phase diagram that is dramatically asymmetrical in $\lambda_2\rightarrow-\lambda_2$. This phase diagram is observed for any (odd) order $n$ of the gain-loss potential, and shows that the threshold can be increased, $\beta_{nc}(\lambda_2)\gg n$ for $\lambda_2\gg1$. In contrast, the bottom row in fig.~\ref{fig:nhh24} shows that the $\PT$ phase diagram is symmetric in $\lambda_4\rightarrow-\lambda_4$ for potential $U_4$. It also shows that $\PT$ threshold is maximum at $\lambda_4=0$ and decreases  monotonically with the strength of the Hermitian potential, $\beta_{nc}(\lambda_4)\leq n$. This symmetric phase diagram is observed for any (odd) order $n$ of the gain-loss potential. The phase diagrams for potential $U_6$, with four different gain-loss potentials $V_n$, are shown in fig.~\ref{fig:nhh6}. The left-hand column shows that for $n=\{1,7\}$, the phase diagram is symmetric in $\lambda_6\rightarrow-\lambda_6$, and the threshold is bounded by its value at the origin, $\beta_{nc}(\lambda_6)\leq n$.  The right-hand column shows that for $n=\{3,9\}$, the $\PT$ threshold increases for positive $\lambda_6$ and is suppressed for $\lambda_6<0$. 

In general, we find that when the order of the $\PT$ potential $n$ is a multiple of $p/2$, where $p$ is the order of the Hermitian potential, the $\PT$ threshold is strengthened for positive $\lambda_p$ and suppressed for $\lambda_p<0$, thus leading to an asymmetrical phase diagram. Otherwise, the $\PT$ phase diagram is symmetric in $\pm\lambda_p$ and the threshold is monotonically suppressed from its $\lambda_p=0$ value, $|\beta_{nc}(\lambda_p)|\leq\beta_{nA}=n$. 
These two distinct  trends in $\PT$ phase diagrams can be qualitatively understood at small $\lambda$ as follows. In the presence of potential $U_p$ eq.(\ref{eq:A}), which determines the $\PT$ threshold, changes to $A'_{mm'}=A_{mm'}-\lambda\left(\delta_{m,m'+p}+\delta_{m,m'-p}\right)/2.$ For all levels $m$, the energy shift due to $U_p$ is quadratic in $\lambda_p$ which, then, reflects in a threshold $\beta_{nc}$ that is even in $\lambda_p$. The sole exception is the pair of degenerate levels $m=\pm p/2$, whose energy shifts are linear in $\lambda_p$,  leading to asymmetric threshold behavior. However, a rigorous proof of numerically obtained symmetries in the threshold $\beta_{nc}(\lambda_p)$ under $\lambda_p\leftrightarrow-\lambda_p$ in figs.(\ref{fig:nhh24})-(\ref{fig:nhh6}) for arbitrary potential strengths is an open question. 


\section{Conclusion}
In this paper, we have investigated a 2D $\PT$-symmetric, continuum model through a family of analytically tractable gain-loss potentials. This model is applicable, in the paraxial approximation, to mode propagation in a single waveguide with annular or circular cross section, and modulated gain and loss. Our results, particularly those in fig.(\ref{fig:nhh24}) and fig.(\ref{fig:nhh6}), have shown that the $\PT$ phase diagrams in the presence of competing potentials, Hermitian or purely gain-loss, offer an unprecedented ability of tune the $\PT$ transitions in such systems. 


\acknowledgments
This work is supported by NSF DMR-1054020 (YJ). RKP thanks the Center for Development of Advanced Computing (C-DAC) for computing time, and KSA thanks the S.P. Pune University for a graduate fellowship. 



\end{document}